\documentclass[manuscript]{revtex4}
\usepackage{amsmath}
\usepackage{amssymb}
\usepackage{color}
\usepackage{subfigure}
\usepackage{graphicx}
\begin{document}
\title{Anomalous Photon-Assisted Tunneling in Graphene}
\author{Andrii Iurov$^1$, Godfrey Gumbs$^{1}$, Oleksiy Roslyak$^{1}$ and Danhong Huang$^{2}$}
\address{$^1$ Department of Physics and Astronomy, Hunter College,
City University of New York; 695 Park Ave., New York, NY 10065, USA}
\address{$^2$ Air Force Research laboratory (ARFL/RVSS), Kirtland Air
Force Base, NM 87117, USA}

\begin{abstract}
We investigated the Dirac electrons
transmission through a potential barrier in the presence of
circularly polarized light. An anomalous photon-assisted enhanced
transmission is predicted and explained in a comparison with the
well-known Klein paradox. It is demonstrated that the perfect
transmission for nearly-head-on collision in an infinite graphene is
suppressed in gapped dressed states of electrons, which is further
accompanied by shift of peaks as a function of the incident angle
away from the head-on collision. In addition, the perfect
transmission in the absence of potential barrier is partially
suppressed by a photon-induced gap in illuminated graphene. After
the effect of rough edges of the potential barrier or impurity
scattering is included, the perfect transmission with no potential
barrier becomes completely suppressed and the energy range for the
photon-assisted perfect transmission is reduced at the same time.
\end{abstract}

\maketitle

\section{Introduction}
\label{sec1}

Research activity on the properties of graphene has been  growing
rapidly  ever since  its experimental discovery  and demonstration
of  its unusual properties arising from its energy band
structure\,\cite{novoselov-main,geim:2009}.  The  novel properties
of graphene may be attributed to its massless Dirac fermions at the
Fermi energy\,\cite{castro-neto}. An interesting  consequence of the
Dirac  electron is  the Klein paradox\,\cite{Katsnelson} in which an
electron in graphene undergoes unimpeded tunneling through potential
barriers of arbitrary height and thickness. This property of Dirac
electrons is due to their linear energy dispersion relation or
helicity. Electrons are said to be chiral if their
wave functions are eigenstates of the chirality operator
$\hat{h}=\mathbf{\sigma}\cdot \mathbf{p}/(2p)$ where
$\mathbf{\sigma}=\{\sigma_x,\,\sigma_y\}$ is the Pauli vector
consisting of Pauli matrices and $\mathbf{p}=\{p_x,\,p_y\}$ is the
electron momentum in graphene layers. Electrons in
graphene near the $K$ points (around the corners of the
hexagonal Brillouin zone) are chiral due to the fact
that the chirality operator is  proportional to the Dirac
Hamiltonian which automatically makes chirality a good quantum
number.
\medskip

The fact that the Klein paradox is also obtained in bilayer graphene
makes this effect even more sophisticated. This leads us to realize
that the Klein paradox is not simply due to linear electron
dispersion but may be observed for both massless and
massive quasiparticles\,\cite{Katsnelson}. In this paper, we
consider a sharp  $p-n-p$ junction or  potential barrier
profile. This type of potential can be
constructed by underlying metal contact or insulating
strip\,\cite{pnp} and was employed to demonstrate unimpeded
transmission\,\cite{castro-neto}.
\medskip

Both the tight-binding model and  ${\bf
k}\cdot{\bf p}$ approximations for infinite graphene
sheet accurately show that electrons and holes have
linear energy dispersions $\varepsilon(k)=s\hbar
v_F\,k=s\hbar v_F \sqrt{k_x^2+k_y^2}$ with no gap, where $s$ is
the electron-hole parity with $s=1$ for electrons and $s=-1$ for
holes. For the potential barrier described above, there is a
translational symmetry in the $y$ direction parallel to the
boundaries of the potential  so that $k_y$ is conserved. In
contrast, the longitudinal component $k_x$ is modified by the
potential so that when the particle has energy $E$, we have
$k_{xi}=\sqrt{(E-V_{i})^2/(\hbar v_F)^2-k_y^2}$,
where $V_i$ is the potential in the region $i$.
\medskip

The electron effective  mass and its properties will play a crucial
role in our analysis. In infinite intrinsic graphene, the Dirac
electron is  massless although the mass can be
introduced\,\cite{Barb1} or implemented experimentally. For  bilayer
graphene, the particle effective  mass exists in any possible
approximation\,\cite{falko-bilayer}, leading to the existence of
evanescent terms in the wave function\,\cite{Katsnelson} although
this does not violate chirality symmetry and the Klein paradox.
\medskip

The outline of the rest of this paper is as follows. In Sec.\
\ref{sec2}, we derive the  eigenstates and transmission coefficient
for Dirac electrons dressed with photons from a circularly polarized
light. Section \ref{sec3} is devoted to a formalism in which the
roughness of the boundaries for the potential barrier is included
phenomenologically in calculating the transmission coefficient and
numerical results are presented.   A brief summary is given in Sec.\
\ref{sec4}.

\section{Transmission Coefficient for Dressed Electron States }
\label{sec2}

It was shown recently\,\cite{Kibis,kibis2} that when Dirac electrons
in a single graphene layer are interacting with an
intense circularly polarized light, electron states
will be \textit{dressed} by photons. The main idea
of the present study is to investigate the transmission properties
of such dressed electrons for the case of single
layer graphene. We go beyond the approximations used
in Ref.\,\cite{Kibis} by retaining the results up to
the order of ${\cal O}(\Delta^4)$ so that we are
able to investigate the difference between the dressed states and
massive Dirac electrons described below by the Hamiltonian in
\eqref{sigma3ham}. Here, $\Delta$ is a
quantity measuring the induced gap between the
valence and conduction bands of dressed electrons.
\medskip

We begin with the electron-photon interaction Hamiltonian

\begin{equation}
\label{initial-ham}
\mathcal{H}=v_F\,\sigma \cdot \left({\mathbf{p}-e\,\mathbf{A}_{circ}}\right) \ ,
\end{equation}
where $v_F$ is the Fermi velocity and the vector potential for
circularly polarized light of frequency $\omega_0$
can be expressed as

\begin{equation}
\mathbf{A}_{circ}=\sqrt{\frac{\hbar}{\epsilon_0\,\omega_0 V}} \left({ \mathbf{e}_+ \hat{a} + \mathbf{e}_- \hat{a}^{\dag} }\right)=\sqrt{\frac{\hbar}{2\epsilon_0\,\omega_0 V}} \left[{ (\hat{a}+\hat{a}^{\dag}) \mathbf{e}_x +i (\hat{a}-\hat{a}^{\dag}) \mathbf{e}_y }\right]
\end{equation}
in terms of photon creation and destruction operators
$\hat{a}^{\dag}$ and $\hat{a}^{\dag}$, respectively.
Here, $V$ is the mode volume of an optical field. In
order to study the complete electron-photon interacting
system, we must add the field energy term $\hbar \omega_0\,
\hat{a}^{\dagger} \hat{a}$ to the Hamiltonian \eqref{initial-ham}.
As usual, we seek the wave function in the form of a plane wave
$\Psi(\mathbf{r})=e^{i\mathbf{k}\cdot \mathbf{r}} \psi(k)$, which
results in the following reduced Hamiltonian

\begin{equation}
\label{mainham}
\mathcal{H}=\hbar\omega_0\,\hat{a}^{\dag} \hat{a} + \hbar v_F\sigma\cdot\mathbf{k}
- \sqrt{\frac{2\hbar e^2v_F^2}{\epsilon_0\,\omega_0 V}} \left({\sigma_+\hat{a}
+\sigma_-\hat{a}^{\dag}}\right) \ ,
\end{equation}
where $\sigma_{+}=\frac{1}{2}
\left({\sigma_x+i\sigma_y}\right)=\left[{ \begin{array}{cc} 0 & 1 \\
0 & 0 \end{array} }\right]$ and $\sigma_{-}=\frac{1}{2}
\left({\sigma_x-i\sigma_y}\right)=\left[{ \begin{array}{cc} 0 & 0
\\1 & 0 \end{array} }\right]$. The reduced Hamiltonian
\eqref{mainham} for infinite graphene can be
understood as  the sum of two parts, namely,  the  Dirac Hamiltonian

\begin{equation}
{\cal H}_{Dirac}=\hbar v_F\,\sigma\cdot \mathbf{k}= \hbar v_F (\sigma_x k_x+\sigma_y k_y)=\hbar v_F (\sigma_- k_+ +\sigma_+ k_-)
\end{equation}
and the Jaynes-Cummings Hamiltonian

\begin{equation}
\label{jcham}
\mathcal{H}_{J-C}=\hbar \omega_0\,\hat{a}^{\dag} \hat{a} -\frac{w}{2\sqrt{N}} \left({ \sigma_+ \hat{a} + \sigma_- \hat{a}^{\dag} }\right)\ ,
\end{equation}
which corresponds to a two-level quantum optical system and, most
importantly, can be solved analytically. Here, we have
defined $k_{\pm}=k_x\pm ik_y$ and $N$ represents the number of
radiation quanta (intensity) for the incident optical field. We
only consider the situation such that the
electron-photon  interaction amplitude $w$ is much less than either
the photon or Dirac electron energy, i.e.,

\begin{equation}
\label{alpha small}
w=2\sqrt{\frac{2N\hbar e^2v_F^2}{\epsilon_0\,\omega_0 V}}=2\alpha\,\hbar\omega_0\ll \hbar \omega_0  \ ,
\end{equation}
where $\alpha\equiv w/\hbar\omega_0\ll 1$. The two
eigenstates of the Hamiltonian \eqref{jcham} could be obtained with
an expansion over the basis of just two functions $\vert 1 \rangle_N
\equiv \vert \uparrow, N \rangle$ and $\vert 2 \rangle_N \equiv
\vert \downarrow, N+1 \rangle $ for each $N$ value, that is,

\begin{gather}
\label{jcfunc1}
\vert \Psi_{\uparrow, N} \rangle = \mu_{N} \vert 1\rangle_N +\nu_{N} \vert 2 \rangle_N\ , \\
\label{jcfunc2} \vert \Psi_{\downarrow, N} \rangle =
\mu_{N} \vert 2 \rangle_N -\nu_{N} \vert 1 \rangle_N\ ,
\end{gather}
which corresponds to the following Hamiltonian in the chosen basis
$\vert 1 \rangle_N$ and $\vert 2 \rangle_N$

\begin{equation}
\mathcal{H}_{N}=\left[{\begin{array}{cc} h_{11} & h_{12} \\
                                       h_{21} & h_{22}

                            \end{array}}\right]= \hbar \omega_0 \left[{\begin{array}{cc} N & \mp\frac{\alpha}{2}\,\sqrt{\frac{N+1}{N}} \\
                                                                                         \mp\frac{\alpha}{2}\,\sqrt{\frac{N+1}{N}} & N+1
                                                                                                \end{array}}\right]\ .
																																																\end{equation}
In this way, the transformation \eqref{jcfunc1} and
\eqref{jcfunc2} becomes just a simple rotation
in the Hilbert space. This expansion yields the
eigenvalue equation and its solution below

\begin{gather}
\left({\varepsilon - N \hbar \omega_0}\right)
\left[{\varepsilon - (N+1) \hbar \omega_0}\right]
- \left({\frac{\alpha}{2}\,\hbar\omega_0\,\sqrt{\frac{N+1}{N}}}\right)^2 = 0\ , \\
\label{jcenergy}
\varepsilon_{\uparrow \downarrow} = \left({N+\frac{1}{2}}\right) \hbar \omega_0 \mp
\frac{\hbar \omega_0}{2} \sqrt{1+\alpha^2 \left({\frac{N+1}{N}}\right)} \backsimeq  \left(N+\frac{1}{2} \mp \frac{1}{2}\right)\hbar \omega_0 \mp\frac{\alpha^2}{4}\hbar \omega_0 \ ,
\end{gather}
where $\epsilon_{\uparrow}$ and
$\epsilon_{\downarrow}$ correspond to the lower $+$ and upper $-$
signs in the solution. For simplicity, we assume here the
radiation is classically strong with $N \gg 1$. However, we note
that we may set $N+1 \backsimeq N$ only in the terms
${\cal O}(\alpha^2)$ but not in the terms with
$N\hbar \omega_0$. Furthermore, it is a simple matter to  obtain the
expansion coefficients $\mu_N$ and $\nu_N$ as

\begin{eqnarray}
\mu_N = \cos\theta_c \backsimeq 1-\frac{\alpha^2}{8}\ , 
\nonumber\\
\nu_N = \sin\theta_c \backsimeq \frac{\alpha}{2}\ ,
\nonumber\\
\tan\theta_c=\frac{\alpha\,\sqrt{(N+1)/N}}{1+\sqrt{1+\alpha^2(N+1)/N}}
\backsimeq \frac{\alpha}{2}\ .
\end{eqnarray}
One can easily verify that the wave functions
\eqref{jcfunc1} and \eqref{jcfunc2} are the eigenstates of the
Jaynes-Cummings Hamiltonian \eqref{jcham} with energies
\eqref{jcenergy}, to any order of $\alpha$. Here, for
all the above derivations, we have employed the standard relations,
i.e.,

\begin{eqnarray}
\label{basis1}
\hat{a}^{\dag} \vert \uparrow \downarrow, N \rangle
&=& \sqrt{N+1} \vert \uparrow \downarrow, N+1 \rangle\ , \hspace{0.3 in}
\hat{a} \vert \uparrow \downarrow, N \rangle=\sqrt{N} \vert \uparrow \downarrow, N-1 \rangle\ ,
\nonumber\\
\sigma_{+} \vert \downarrow, N \rangle
&=& \vert \uparrow, N \rangle\ , \hspace{0.5 in} \sigma_+ \vert \uparrow, N \rangle = 0\ ,
\nonumber\\
\sigma_{-} \vert \uparrow, N \rangle
&=& \vert \downarrow, N \rangle\ , \hspace{0.5 in} \sigma_- \vert \downarrow ,N \rangle = 0 \ .
\end{eqnarray}
We now look for the eigenstates of the Hamiltonian
\eqref{mainham} as expansions over the set of Jaynes-Cummings
Hamiltonian eigenfunctions \eqref{jcfunc1} and \eqref{jcfunc2}. We
will confine our attention to  the field source with only three
nearest photon occupation numbers, i.e.,  $N=N_0-1$, $N_0$ and
$N_0+1$, leading to

\begin{equation}
\label{expansion1} \vert \Phi(k) \rangle\backsimeq
\sum_{\ell=N_0-1}^{N_0+1}\left({\mathcal{C}_{1,\;\ell}(k) \mid
\Psi_{\uparrow,\;\ell} \rangle +\mathcal{C}_{2,\;\ell}(k) \mid
\Psi_{\downarrow,\;\ell} \rangle }\right) \ .
\end{equation}
\medskip

We know that the eigenfunctions \eqref{jcfunc1},
\eqref{jcfunc2} corresponding to different numbers $N$ are
orthogonal to each other. First acting the Hamiltonian
\eqref{mainham} on \eqref{expansion1}, and then, multiplying both
sides of the expansion \eqref{expansion1} by $\langle
\Psi_{\uparrow, N_0} \vert$, $\langle \Psi_{\downarrow, N_0-1}
\vert$, $\langle \Psi_{\uparrow, N_0+1} \vert$ and $\langle
\Psi_{\downarrow, N_0} \vert$, this result in the following four
equations

\begin{eqnarray}
\label{sys1} &&\mathcal{C}_{1, N_0}\left({N_0
-\frac{\alpha^2}{4}\,\frac{N_0+1}{N_0}}\right)\hbar\omega_0 +\hbar
v_F \; \langle \Psi_{\uparrow, N_0} \vert \sigma \cdot \mathbf{k}
\vert \Phi(k) \rangle =\varepsilon\,\mathcal{C}_{1,
N_0}\ ,
\nonumber\\
&&\mathcal{C}_{2, N_0-1}\left({N_0
+\frac{\alpha^2}{4}\,\frac{N_0}{N_0-1}}\right)\hbar\omega_0 +\hbar
v_F \; \langle \Psi_{\downarrow, N_0-1} \vert \sigma \cdot
\mathbf{k} \vert \Phi(k) \rangle
=\varepsilon\,\mathcal{C}_{2, N_0-1}\ ,
\end{eqnarray}

\begin{eqnarray}
\label{sys2} &&\mathcal{C}_{1, N_0+1}\left[{(N_0+1)
-\frac{\alpha^2}{4}\,\frac{N_0+2}{N_0+1}}\right]\hbar\omega_0 +\hbar
v_F \; \langle \Psi_{\uparrow, N_0+1} \vert \sigma \cdot \mathbf{k}
\vert \Phi(k) \rangle =\varepsilon\,\mathcal{C}_{1,
N_0+1}\ ,
\nonumber\\
&&\mathcal{C}_{2, N_0}\left[{(N_0+1)
+\frac{\alpha^2}{4}\,\frac{N_0+1}{N_0}}\right]\hbar\omega_0 +\hbar
v_F \; \langle \Psi_{\downarrow, N_0} \vert \sigma \cdot \mathbf{k}
\vert \Phi(k) \rangle =\varepsilon\,\mathcal{C}_{2, N_0}
\ .
\end{eqnarray}
For chosen number $N_0$, each pair of these equations  describe two
energy subbands that are separated  by $ \backsimeq
\alpha^2\hbar \omega_0/2$ at $k=0$. Here, we include only four
nearest energy subbands, corresponding to the dressed states with
different photon occupation numbers $N$ and electron states with
subband indices $\uparrow$ (lower energy) and
$\downarrow$ (higher energy).
\medskip

Taking into account the following relations for the Dirac Hamiltonian,

\begin{eqnarray}
&&\sigma\cdot \mathbf{k} \; \vert{\Psi_{\uparrow, N_0}}
\rangle = \mu_{N_0} k_+ \vert \downarrow, N_0 \rangle - \nu_{N_0}
k_- \vert \uparrow, N_0+1 \rangle\ ,
\nonumber\\
&&\sigma\cdot \mathbf{k} \; \vert{\Psi_{\downarrow,
N_0}} \rangle = -\mu_{N_0} k_- \vert \uparrow, N_0+1 \rangle -
\nu_{N_0} k_+ \vert \downarrow, N_0 \rangle\ ,
\end{eqnarray}
and with the simplifications described above, we can
explicitly write out the Dirac Hamiltonian terms in equations
\eqref{sys1} and \eqref{sys2}. The presence of photon occupation
numbers does not follow from the dressed states Hamiltonian and
cannot be determined only from the ratio between the photon filed
energy $N \hbar \omega_0$ and the electron-photon interaction
amplitude $w$. It should results from the model of the circularly
polarized light source. If we consider electrons near
the $K$ point such that the kinetic energy $\hbar v_F\,k \backsimeq
\alpha^2 \hbar \omega_0/2$, we can only retain only three photon
occupation numbers $N_0-1$, $N_0$ and $N_0+1$, which is consistent
with the approximation described above. Consequently, we arrive at
the system which consists of weakly coupled equations to determine
the two nearest subbands, i.e.,

\begin{equation}
\left[{\begin{array}{cc|cc}
N_0 \hbar \omega_0 - \Delta & -\mu^2 \hbar v_Fk_{-}  & \mu\nu \hbar v_Fk_{+}  & 0 \\
\mu^2 \hbar v_Fk_{+} & N_0 \hbar \omega_0 + \Delta  & 0 & -\nu \mu \hbar v_Fk_{+}\\ \hline
-\nu \mu \hbar v_Fk_{-} & 0  & (N_0+1) \hbar \omega_0 - \Delta & -\mu^2 \hbar v_Fk_{-} \\
0 & \nu \mu \hbar v_Fk_{-} & \mu^2 \hbar v_Fk_{+} & (N_0+1) \hbar
\omega_0 + \Delta  \end{array} }\right]\, \left[{\begin{array}{c}
C_1
\\ C_2 \\ C_3  \\ C_4
\end{array}}\right]=\varepsilon\,\left[{\begin{array}{c} C_1 \\ C_2 \\ C_3  \\  C_4 \end{array}}\right]\ ,
\end{equation}
where $\mu\equiv\mu_{N_0}$, $\nu\equiv\nu_{N_0}$,
$\Delta \backsimeq \alpha^2 \hbar \omega_0/4$, $C_1\equiv
C_{1,N_0}$, $C_2\equiv C_{2,N_0-1}$, $C_3\equiv C_{1,N_0+1}$ and
$C_4\equiv C_{2,N_0}$. This  leads to the following energy
dispersions with $\mu\approx 1$:

\begin{gather}
\varepsilon_{1,\,2}(k)={\left(N_0+\frac{1}{2}\right)\hbar
\omega_0 -\frac{1}{2}\,\hbar\omega_0\sqrt{1+\eta+\xi(1+\nu^2)k^2
\pm 2\sqrt{(\xi\nu^2 k^2+1)(\xi k^2+\eta)}}  }\ , \\
\varepsilon_{3,\,4}(k)={\left(N_0+\frac{1}{2}\right)\hbar
\omega_0 +\frac{1}{2}\,\hbar\omega_0\sqrt{1+\eta+\xi(1+\nu^2)k^2 \mp
2\sqrt{(\xi\nu^2 k^2+1)(\xi k^2+\eta)}}  } \ ,
\end{gather}
where $\xi=(2v_F/\omega_0)^2$ and
$\eta=(2\Delta/\hbar\omega_0)^2\approx\alpha^4/4$. Another
approximation which we may use here is to consider only small wave
vectors $k$ so that we get two independent pairs of subbands
separated by energy $\hbar \omega_0$. This gives a
two-component spinor wave function where each component consists of
two independent terms. If we consider the transmission of such
states through a potential barrier whose height is equal to $\hbar
\omega_0$, one of the terms in region ``1'' will exactly
match the other in the region ``2'' (the potential
region) so that a part of the wave function will be completely
transmitted. This will lead to a substantial increase in the total
transmission. On the other hand, if we consider a
larger number of such subband pairs, this effect will not be
significant since this complete transmission occurs only for one of
the wave function terms. This is the reminiscent of
the Klein paradox since it does not depend on the barrier width as
long as the barrier height is exactly equal to $\hbar \omega_0$.
Revealing such an anomalous increase for the dressed states
tunneling is an important discovery of the present
paper. The corresponding eigenvalue equation for the case of two
independent pairs of subbands yields, as expected

\begin{equation}
\left\{\hbar^2
v_F^2\,k^2+\Delta^2-\left(\varepsilon-N_0\,\hbar\omega_0\right)^2\right\}\,
\left\{\hbar^2 v_F^2\,k^2+\Delta^2-\left[\varepsilon-(N_0+1)\,\hbar
\omega_0\right]^2\right\}=0\ .
\end{equation}
\medskip

The simplest possible approximation is to keep only two terms with
the coefficients $[\mathcal{C}_1(k),\,\mathcal{C}_2(k)]$ leading to
the two nearest energy subbands separated by a gap
$2\Delta \backsimeq \alpha^2\hbar \omega_0/2$ at
$k=0$ due to electron-photon interaction. This approximation
results in a simplified algebraic system determining the pair of
coefficients $\mathcal{C}_1(k)$ and $\mathcal{C}_2(k)$, i.e.,

\begin{eqnarray}
\notag
(N_0\,\hbar \omega_0 - \Delta)\,\mathcal{C}_1 + \hbar v_F\,(k_x+i k_y)\,\mathcal{C}_2
= \varepsilon\,\mathcal{C}_1\ ,   \\
\label{systema}
(N_0\,\hbar \omega_0 + \Delta)\,\mathcal{C}_2 + \hbar v_F\,(k_x-i k_y)\,\mathcal{C}_1
= \varepsilon\,\mathcal{C}_2 \ .
\end{eqnarray}
The non-trivial solution of these equations gives the energy dispersion which
was previously obtained in Ref.\,\cite{Kibis} as

\begin{equation}
\label{energyk} \varepsilon(k)=N_0\,\hbar \omega_0 \pm
\sqrt{\Delta^2+\hbar^2 v_F^2\,k^2}\equiv N_0\,\hbar \omega_0 + \beta
\sqrt{\Delta^2+\hbar^2 v_F^2\,k^2} \ .
\end{equation}
If the electron-photon interaction is removed, then
$\alpha\equiv w/2 \to 0$, and the energy dispersion
relations \eqref{energyk} demonstrate a non-interacting system
consisting of a Dirac electron $\varepsilon_\beta (k)=
\beta\hbar v_F\,\vert k \vert$ and photons $N_0\,\hbar \omega_0$.
All the other energy subbands are separated at least
by $\hbar \omega_0 \gg w$ and could be neglected, which justifies
the above two-subband approximation. In this
notation, $\beta=\pm 1$ is the dressed
conduction/valence band index corresponding to bare
electron/hole bands for infinite graphene when $\Delta \to 0$.
\medskip

The system \eqref{systema} is formally similar to the eigenvalue equations for the
case of the effective-mass Dirac Hamiltonian

\begin{equation}
\label{sigma3ham} \mathcal{H}=\hbar v_F\,\sigma \cdot \mathbf{k}+
\mathcal{V}(x) \left[{ \begin{array}{cc} 1 & 0 \\ 0 & 1
\end{array} }\right] +\Delta\,\sigma_{3} \ ,
\end{equation}
where $\sigma_{3}$ is a Pauli matrix and $\mathcal{V}(x)$ is a
one-dimensional potential. The electron dispersion and transmission
properties for  both a single as well as multiple square potential
barriers have been studied\,\cite{Barb1, Barb2} for monolayer and
bilayer graphene\,\cite{Barb3}. It was also shown that a
one-dimensional periodic array of potential barriers  leads to
multiple  Dirac points\,\cite{Barb4}. Several papers have introduced
an effective mass term into the Dirac Hamiltonian for infinite
graphene  which may be justified based on  different physical
reasons\,\cite{gapstun}. For example, it has been
shown\,\cite{substrate} that an energy bandgap in graphene can be
created by boron nitride substrate resulting in a
finite electron effective mass. However, we emphasize  that the
analogy between the Hamiltonian \eqref{sigma3ham} and that for
irradiated graphene  is not complete since that would correspond to
$\Delta<0$. Although this difference does not result in
any modification of the energy dispersion term containing
$\Delta^2$, it certainly modifies the corresponding wave function.
\medskip

The interaction between Dirac electrons in graphene and a circularly
polarized light has been considered in the classical limit  in
Ref.\,\cite{aoki_oki}. In this limit, a gap  in the Dirac cone opens
up due to nonlinear effects. The dressed state wave function has the
form

\begin{equation}
\Phi_{dr}(k)= \left[{ \begin{array}{c} \mathcal{C}_1(k) \\ \beta \;
\mathcal{C}_2(k)\,\texttt{e}^{i \phi}  \end{array}  }\right]
\end{equation}
with $\mathcal{C}_1(k) \neq \mathcal{C}_2 (k)$ given by

\begin{eqnarray}
\mathcal{C}^{\pm}_1(k)=\frac{1}{\sqrt{2(1+\gamma^2) \mp 2\gamma\sqrt{1+\gamma^2}}}\ , \\
\mathcal{C}^{\pm}_2(k)=\pm
\frac{\sqrt{1+\gamma^2}\mp\gamma}{\sqrt{2(1+\gamma^2) \mp
2\gamma\sqrt{1+\gamma^2}}}\ ,
\end{eqnarray}
corresponding to the energy subbands $\varepsilon(k)=N_0\,\hbar
\omega_0 \pm \sqrt{\Delta^2+\hbar^2 v_F^2\,k^2}$. In this notation,
$\gamma=\Delta/(\hbar v_F\,k)$ and $\phi$ is the angle which
${\mathbf k}$ makes with the longitudinal $x$ axis.
Without an optical field, i.e. $\Delta=0$, we obtain
$\mathcal{C}^{\pm}_1=\mathcal{C}^{\pm}_2=1/\sqrt{2}$. In the limit
$(\Delta \ll k)$, the coefficients exhibit peculiar symmetry with

\begin{eqnarray}
\label{approx-coeff} \mathcal{C}^{+}_{1,2}(k)& \backsimeq &
\frac{1}{\sqrt{2}} \pm \frac{\gamma}{2\sqrt{2}} -
\frac{\gamma^2}{8\sqrt{2}} \mp \frac{3 \gamma^3}{16\sqrt{2}}\ ,
\nonumber\\
\mathcal{C}^{-}_{1,2}(k)&   \backsimeq & \pm \frac{1}{\sqrt{2}} -
\frac{\gamma}{2\sqrt{2}} \mp \frac{\gamma^2}{8\sqrt{2}} + \frac{3
\gamma^3}{16\sqrt{2}}\ .
\end{eqnarray}
We note that this expansion is not valid too close to the Dirac point and should not
be used for arbitrary wave vector to calculate, for example,  the  polarization
function. Additionally, one may  verify that $\mathcal{C}_1(k) \neq \mathcal{C}_2(k)$
for any chosen $\Delta$ in the range of validity. Consequently,
the chiral symmetry is broken for electron dressed states

\begin{equation}
\label{non_chiral} \hat{h}\Psi_{dr}(k) =\frac{1}{2} \,\frac{\sigma
\cdot \mathbf{p}}{p} \left[{ \begin{array}{c} \mathcal{C}_1(k)
\\ \beta \; \mathcal{C}_2(k)\, \texttt{e}^{i \phi}
\end{array}  }\right] = \frac{1}{2}\left[{ \begin{array}{c}
\beta \; \mathcal{C}_2(k) \\ \mathcal{C}_1(k)\, \texttt{e}^{i \phi}
\end{array}  }\right] \  .
\end{equation}
Clearly, it follows from \eqref{non_chiral} that the
non-chirality of the dressed electron states becomes significant if
the electron-photon interaction (the leading $\gamma$ term) is
increased. This affects the electron tunneling and transport
properties. We now turn to an investigation of the transmission of
electron states through a potential barrier when graphene is
irradiated with a circularly polarized light.
\medskip

For simplicity, we consider a square potential  barrier of height
$V_0$ given by ${\cal V}(x)=V_0
\left[{\theta(x)-\theta(x-W_0)}\right]$ where $W_0$ is the barrier
width and $\theta(x)$ is the Heaviside step function.
Since the wave number $k_x$ is the same in region $1$
($x<0$) and region $3$ ($x>W_0$) and the current component is
$j_x=\Phi^{\dag}\sigma_x\Phi$, we only need the wave-function
continuity at the potential boundaries for the system considered.
From this continuity condition, the transmission probability $T$ can
be determined from $T=\vert t \vert^2$ where $t$ is the transmission
coefficient or the amplitude of the wave propagating forward in
region $3$. Here, we only show an analytical expression
for the transmission coefficient of the dressed states in the limit
of $\varepsilon \ll V_0$ corresponding to $k_{x1} \ll k_{x2}$:

\begin{gather}
T=\frac{\cos^2 \phi}{\cos^2{(k_{x2} W_0)\;\cos^2\phi +
\sin^2 (k_{x2} W_0)}} -\frac{3\cos^2(k_{x2} W_0)\,\cos^2\phi\,
\sin^2(k_{x2} W_0)}{\hbar^2v_F^2\,k_{x1}^2 \left[{\sin^2(k_{x2}
W_0)+\cos^2(k_{x2} W_0)\,\cos^2\phi}\right]}\;\Delta^2  \ .
\end{gather}
Here, $\theta=\tan^{-1}(k_y/k_{x2})\to 0$,
$\phi=\tan^{-1}(k_y/k_{x1})$, the second term includes the effect of
electron-photon interaction ($\propto\Delta^2$) to the leading
order. In addition, we only show those terms of the lowest order
in $\varepsilon/V_0$. There is another relevant
study\,\cite{gomes}, investigating the tunneling of Dirac electrons
with a finite effective mass, a parabolic dispersion in the presence
of a energy gap, and a certain chirality, through a potential
region. In that study, the particle tunneling through a square
potential barrier differs from both Dirac electrons and the dressed
states of electrons under a circularly polarized light
illumination.
\medskip

For nearly-head-on collision with $k_y \ll k_{x1}\ll k_{x2}$ for
high potential as well as for infinite graphene ($\Delta\to 0$),
transmission coefficient has the following simplified form

\begin{equation}
T=1-\sin^2(k_{x2}
W_0)\,\left(\theta^2-2\beta\theta\phi+\phi^2\right)\ ,
\end{equation}
where we assume $V_0 \gg \varepsilon$, $\theta \ll \phi \ll 1$ and
$\beta=\pm 1$.

\section{Numerical Results and Discussion}
\label{sec3}

In our numerical calculations, energies will be
measured in units of $(3k_Fat/2)$ with the carbon-carbon distance
$a\approx1.42$\,\AA\, and the hopping parameter $t=2.7
\sqrt{3}/2\;eV$. We measure the wave vector in units of the Fermi
wave number $k_F$ and write its components as $k_x=
\;\cos\phi$ and $k_y= \; \sin\phi$ in terms of the angle of
incidence $\phi$.
\medskip

\begin{figure}[H]
\centering
\subfigure[]{
\includegraphics[height=2.5in]{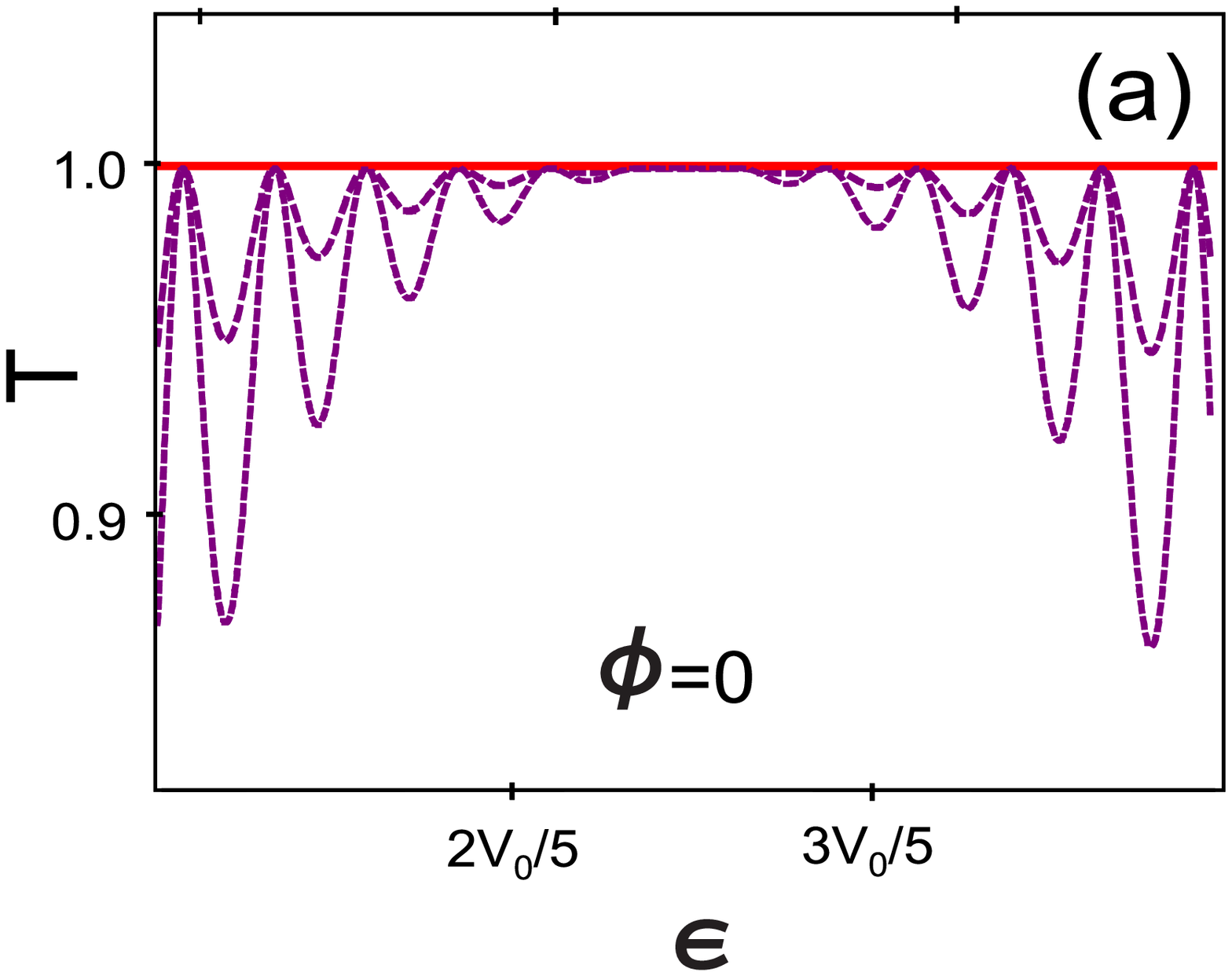}}
\subfigure[]{
\includegraphics[height=2.8in]{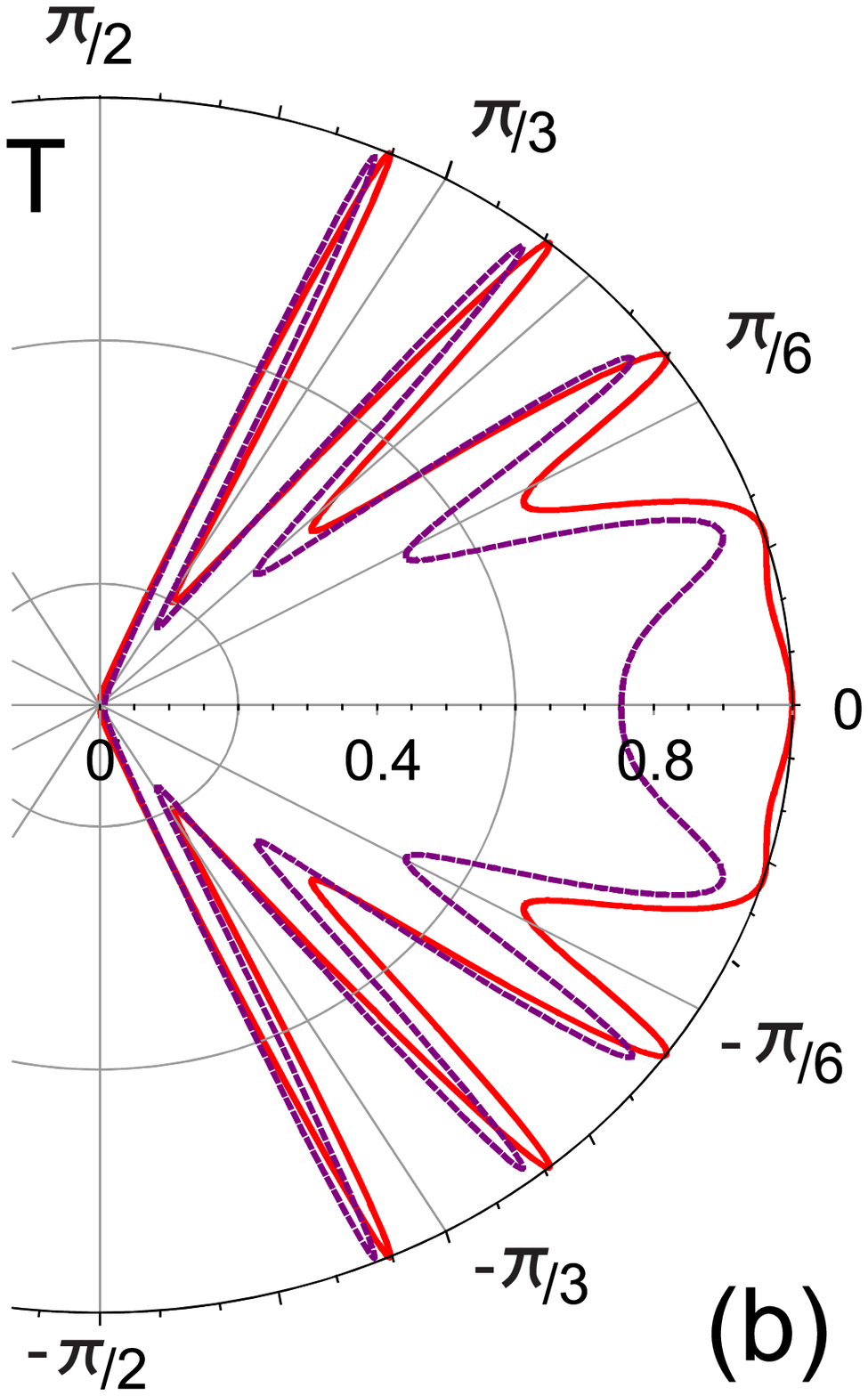}}
\\
\subfigure[]{
\includegraphics[width=0.35\textwidth]{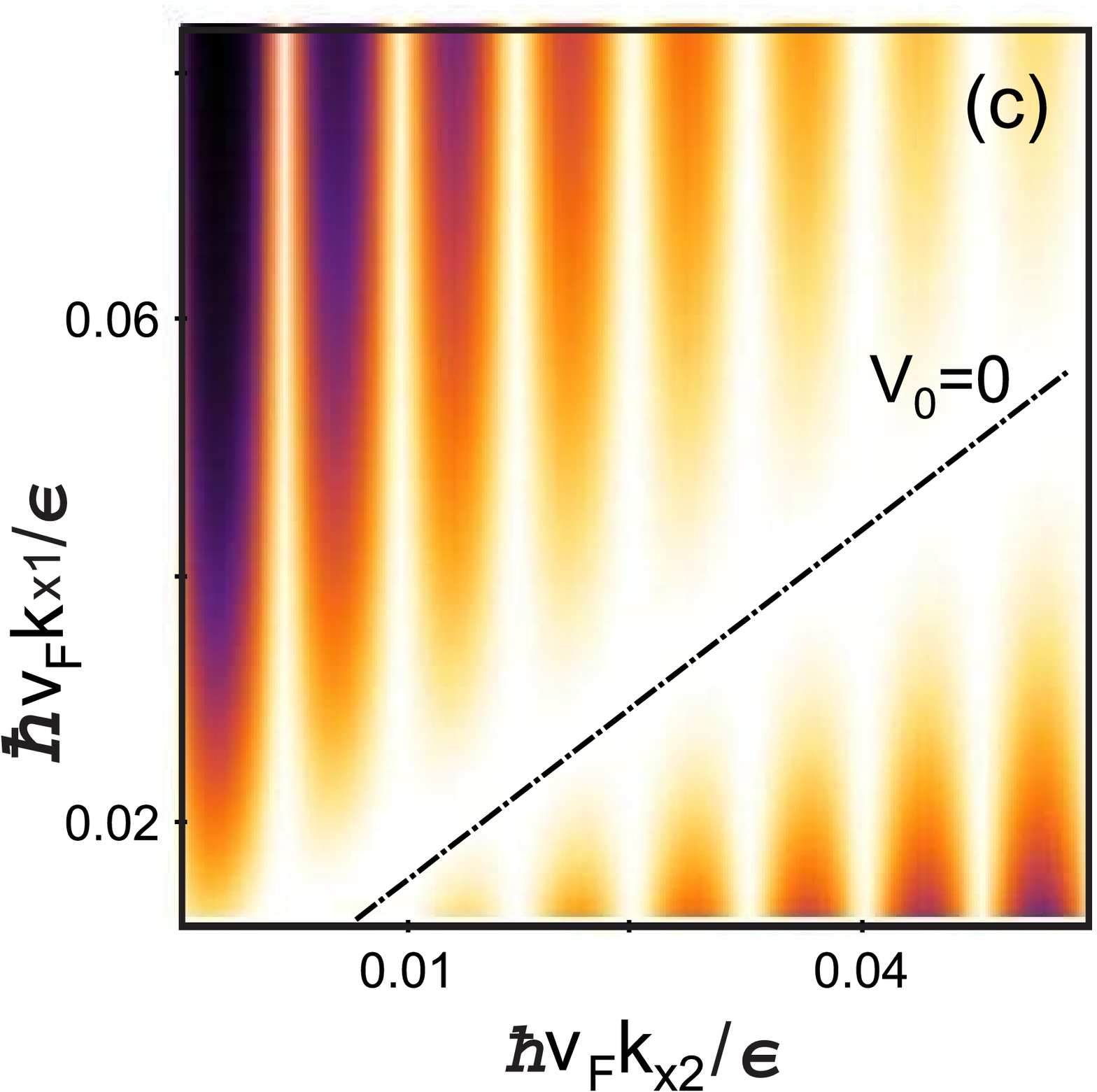}}
\subfigure[]{
\includegraphics[width=0.35\textwidth]{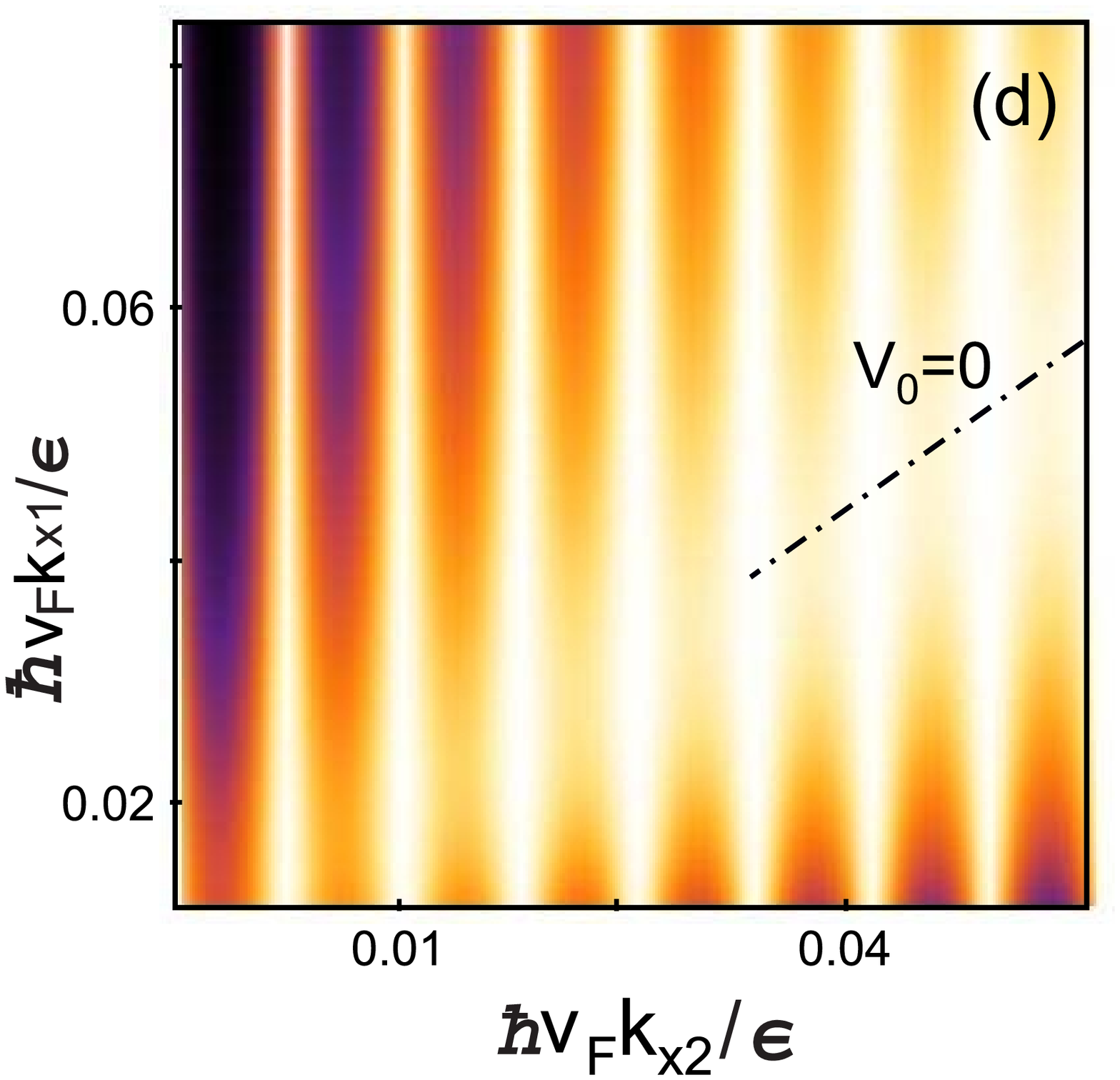}}
\subfigure[]{
\includegraphics[width=0.068\textwidth]{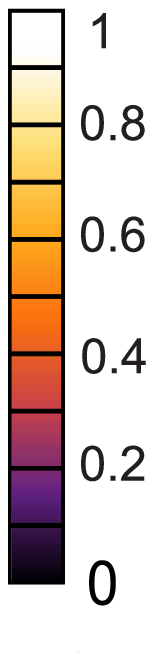}}
\caption{(Color online) Transmission probability $T$ of electrons as
functions of incoming energy $\varepsilon$, incident angle $\phi$,
and longitudinal wave numbers $\hbar v_F\,k_{x1}$ and $\hbar
v_F\,k_{x}$. In (a), the transmission probability (purple dashed
curve) is plotted as a function of the incident particle energy for
head-on collision $(\phi=0)$. The red solid line is the transmission
probability $T=1$ (Klein paradox) for infinite graphene when the
energy gap $\Delta=0$. In (b), we show the angular distributions of
$T$ for infinite graphene (solid red curve) and the dressed states
(purple dashed curve) with $\varepsilon=V_0/6$. In (c) and (d), the
transmission probability is plotted as a function of the
longitudinal momenta in regions ``1'' and ``2'' (before the
potential barrier and in the barrier region, respectively) for
infinite graphene (on the left) and irradiated graphene (on the
right). The diagonal $k_{x1}=k_{x2}$ corresponds to the absence of a
potential barrier ($V_0=0$) and displays a perfect transmission with
$T=1$ as long as both longitudinal momenta are real determined by
both the barrier height $V_0$ and the energy gap $\Delta$.}
\label{F1}
\end{figure}

In Fig.\,\ref{F1}, we present the  transmission for
dressed electron states with arbitrary energy and angle of
incidence. We clearly see that dressing ruins the Klein paradox in
(a) for head-on collision with $\phi=0$. The resonant peaks are
shifted for the other incoming angles in (b) and the effect is
stronger for small incident angles. In (c) and (d), the transmission
probability plots are given in terms of the longitudinal momentum
$k_{x1}$ in front the barrier and $k_{x2}$ in the barrier region. We
find that the intensity and locations of the transmission peaks in
(d) are distorted compared to infinite graphene in (c). The diagonal
$k_{x1}=k_{x2}$ corresponds to the absence of potential barrier and
should yield a complete transmission for $\Delta=0$. However, the
condition $(\sqrt{(\varepsilon-V_0)^2- \Delta^2} > \hbar v_F\,k_y)$
for dressed states must be satisfied, which makes the diagonal
incomplete (missing diagonal for small $k_{x1}$ and $k_{x2}$) due to
the occurrence of an induced gap.
\medskip

\begin{figure}[p]
\centering
\subfigure[]{
\includegraphics[height=1.75in]{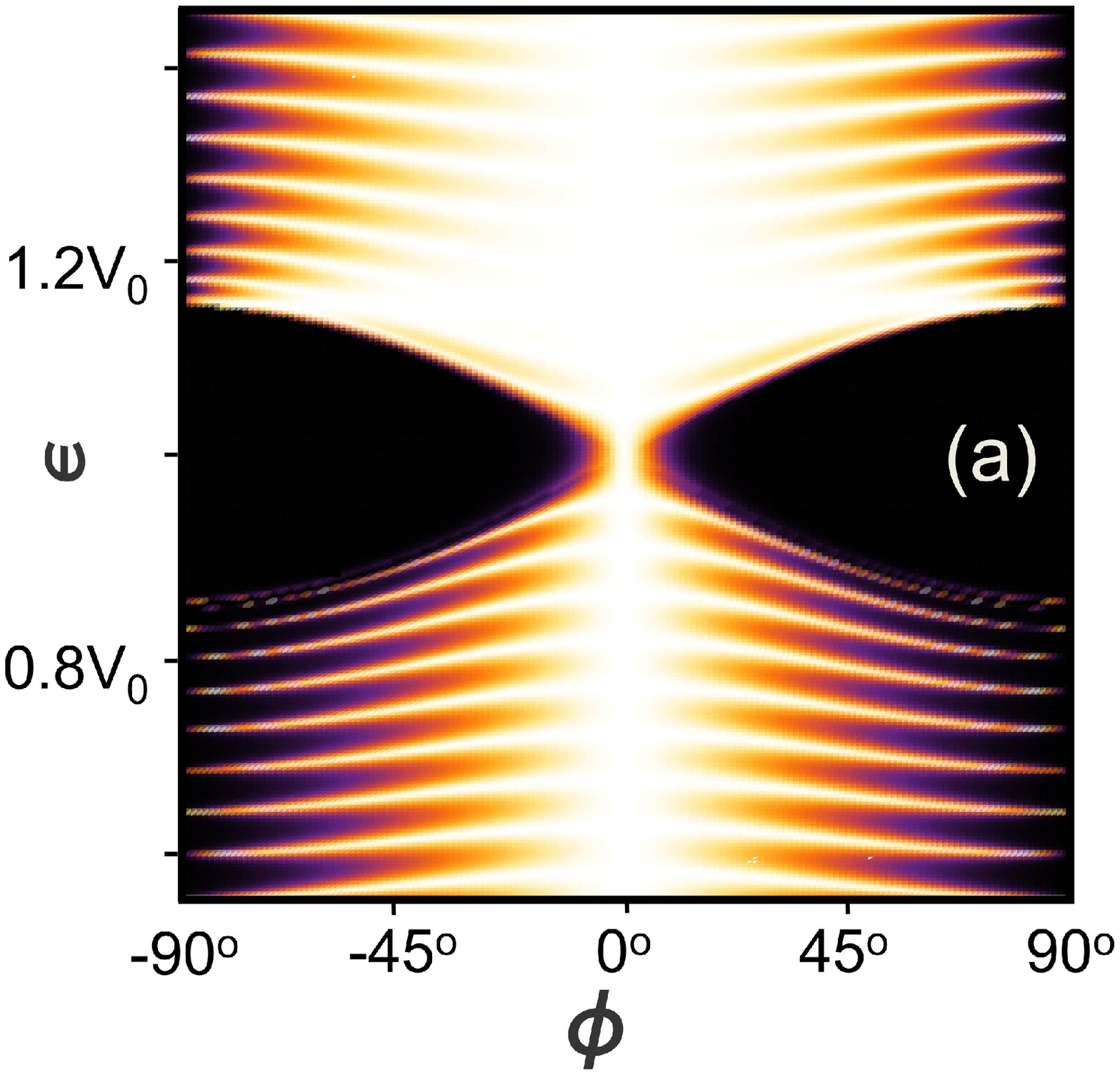}}
\subfigure[]{
\includegraphics[height=1.75in]{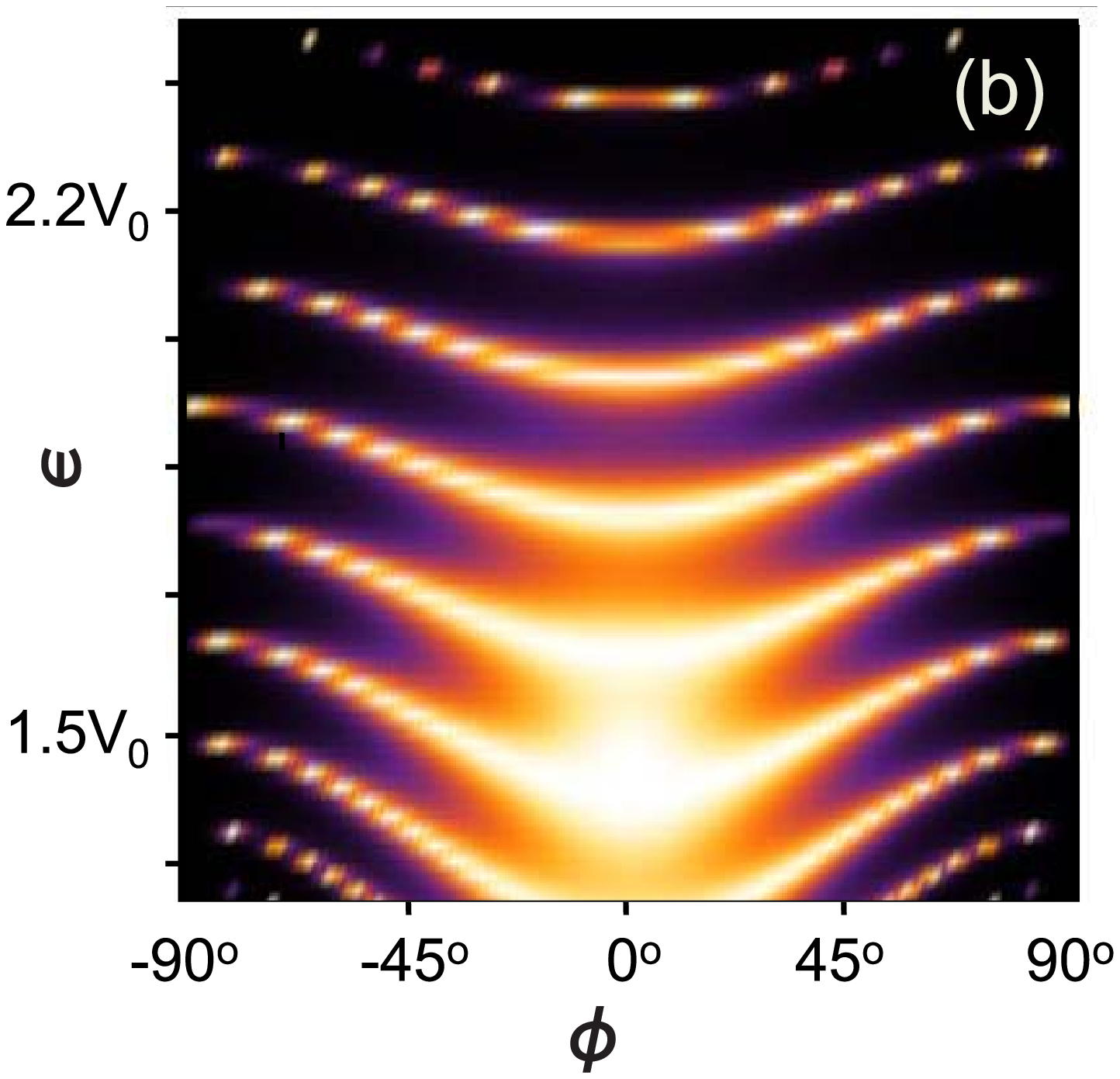}}
\subfigure[]{
\includegraphics[height=1.75in]{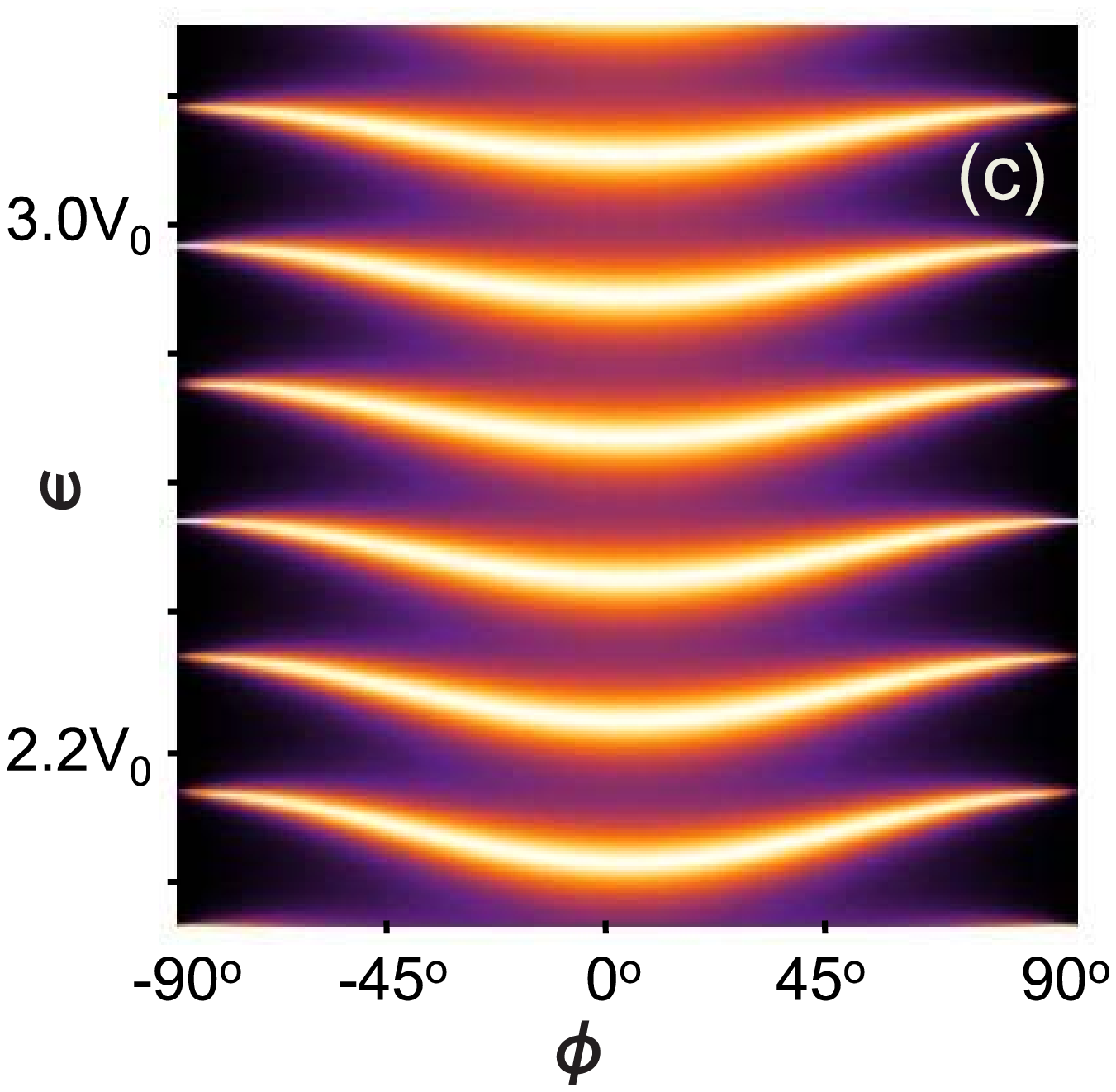}}
\subfigure[]{
\includegraphics[height=1.78in]{colorbar}}
\caption{(Color online) Transmission probability $T$ as functions of
both the incoming electron energy $\varepsilon$ and the angle of
incidence $\phi$. Plot (a) (far left) is for infinite graphene
($\Delta=0$) with obvious Klein paradox for $\phi=0$. Plots (b) and
(c) show the transmission $T$ for electron dressed states, in the
two nearest subbands approximation, with gap energies $\Delta
=V_0/15$ (middle) and $\Delta =V_0/5$ (far right).} \label{F2}
\end{figure}

Figure\ \ref{F2} displays the effect due to
electron-photon interaction on the electron transmission in terms of
incoming particle energy $\varepsilon$ and angle of incidence
$\phi$. From the figure, we see the Klein paradox as well as
resonant tunneling peaks in the transmission probability for regular
infinite graphene with $\Delta=0$ in (a). The dark ``pockets'' on
both sides of $\varepsilon=V_0$ in (a) demonstrate zero transmission
for the case $\vert \varepsilon - V_0 \vert \ll \varepsilon$, which
results in imaginary longitudinal momenta $k_{x2}$ for most of
incident angles and produces a completely attenuating wave function.
When a small gap is opened in (b) for dressed states of electrons in
graphene under the illumination by a circularly-polarized light, we
observe a set of complete transmission branches, where a strong
dependence on $\phi$ for lower branches is seen. However, this
$\phi$ dependence is greatly suppressed when the dressed-state gap
is increased in (c), leaving us a set of equally-distant branches
due to photo-assisted electron tunneling.
\medskip

\begin{figure}[p]
\centering \subfigure[]{
\includegraphics[height=1.7in]{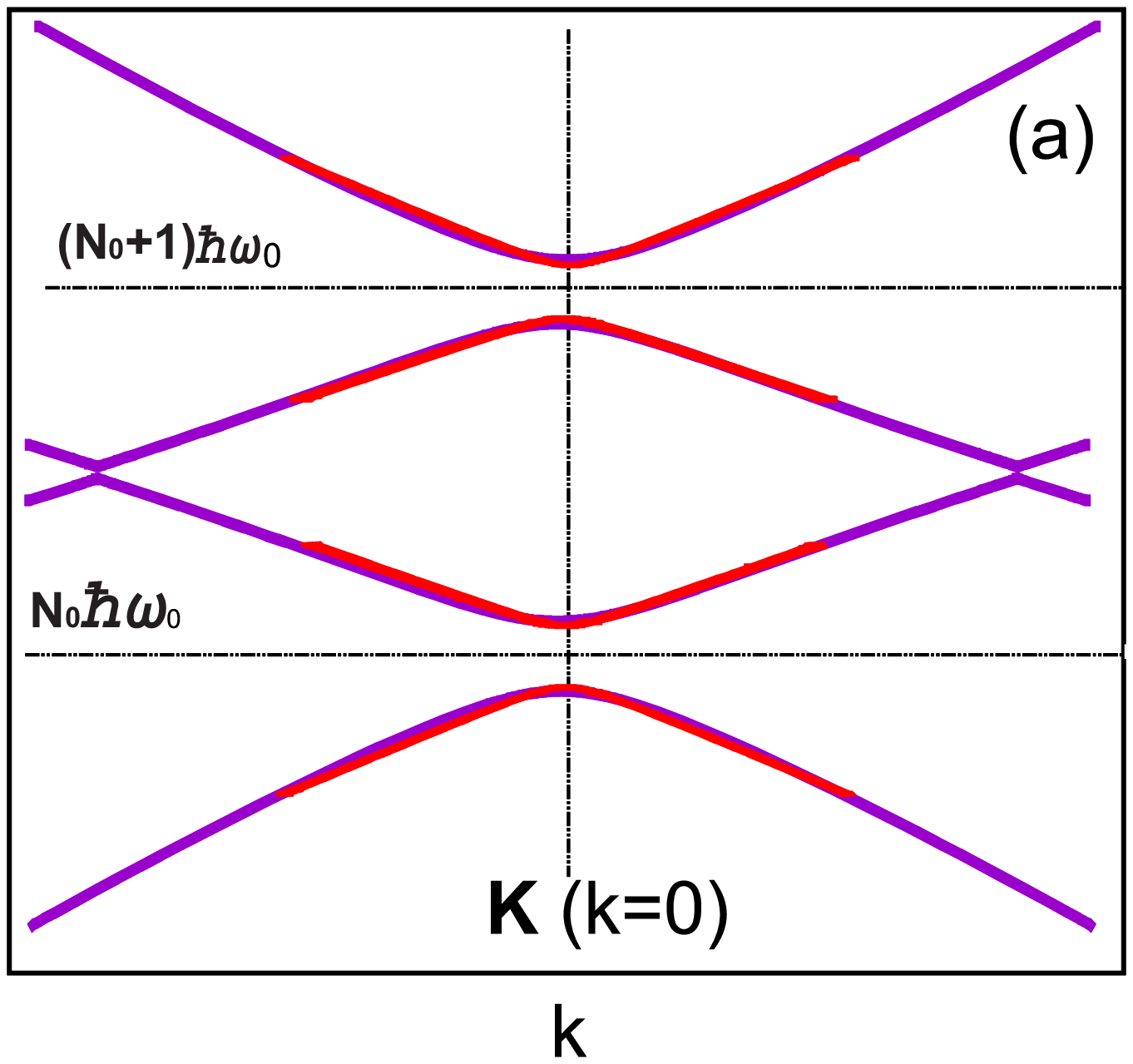}}
\subfigure[]{
\includegraphics[height=1.7in]{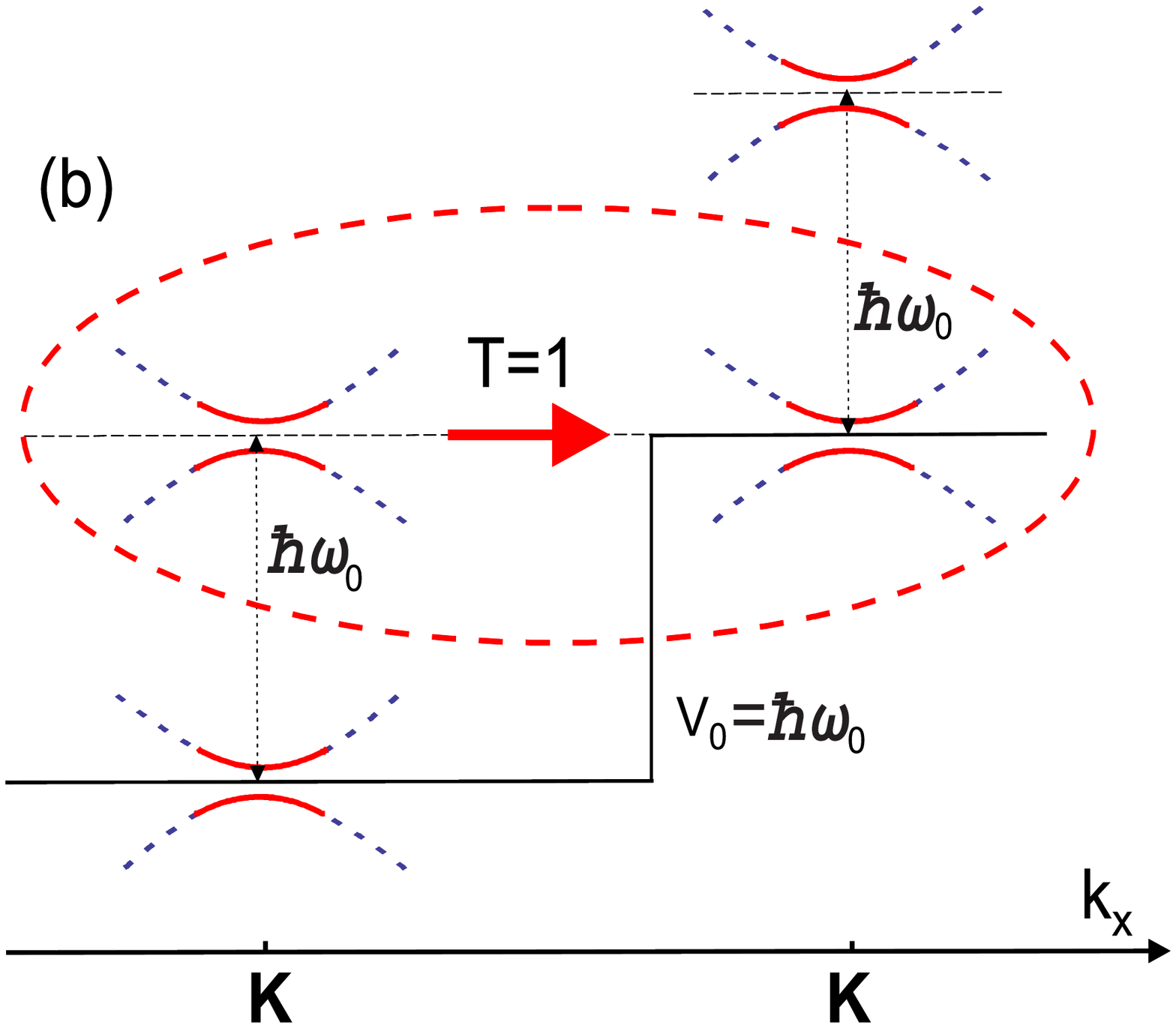}}
\subfigure[]{
\includegraphics[height=1.7in]{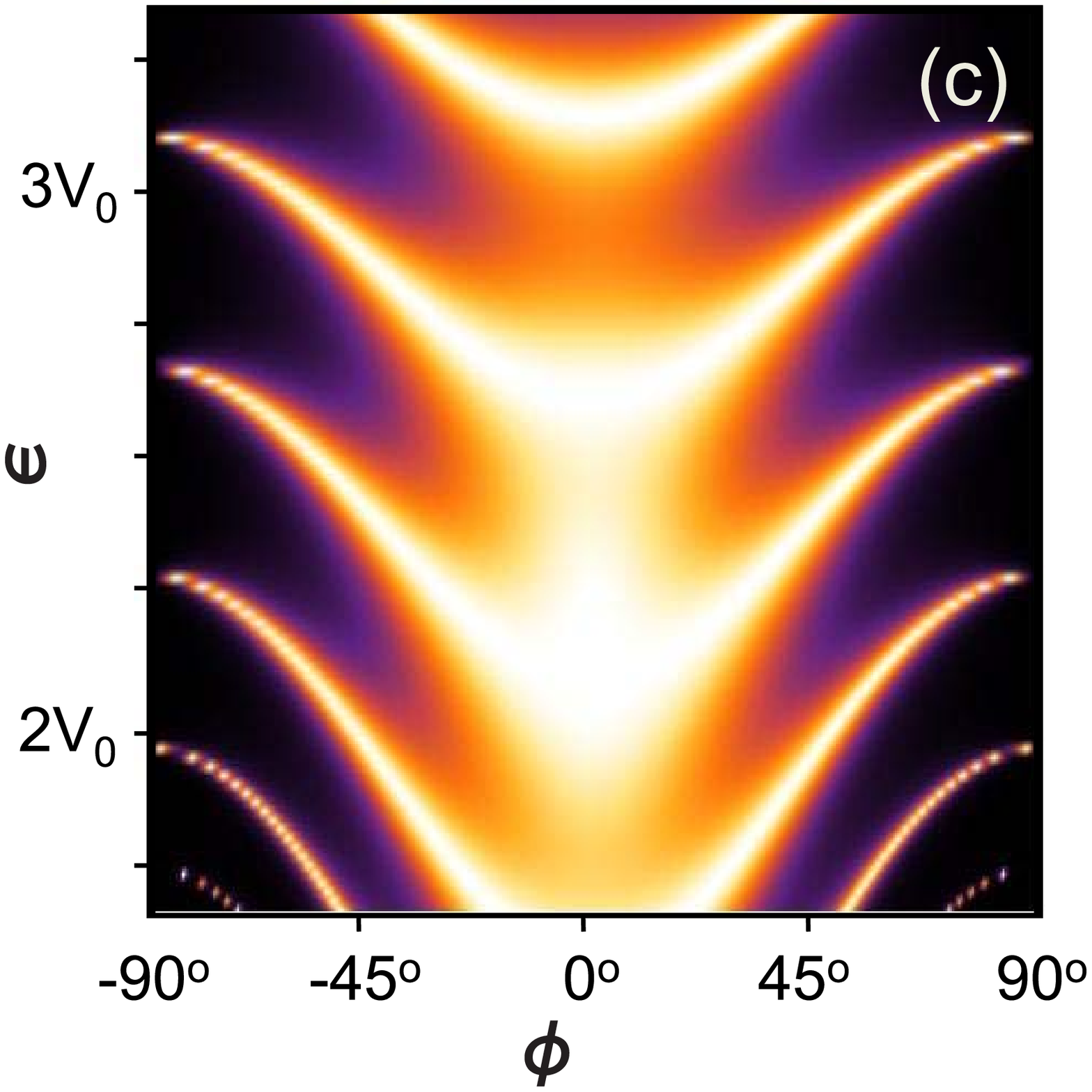}}
\subfigure[]{
\includegraphics[height=1.73in]{colorbar}}
\caption{(Color online)  Energy dispersion and transmission
probability for the case of two independent pairs of energy
subbands. Panel (a) on the left shows the energy dispersion by
taking into account the coupling between the states corresponding to
two different photon occupation numbers in the linear approximation.
In the middle panel (b), we present a  schematic diagram  showing
how the transmission probability may be increased due to the
presence of the other energy subbands. Panel (c) on the right gives
the transmission vs. incoming particle energy and the angle of
incidence for the case of two independent pairs of energy subbands.}
\label{F3}
\end{figure}

As mentioned above, by including  more than the two
nearest subbands, as shown in in Fig.\,\ref{F3}(a), the electron
dispersion, wave function and transmission amplitude will be
modified. According to the approximation adopted by
Kibis\,\cite{Kibis}, the two subband pairs may be considered as
independent. Under the condition of equal transverse momentum $k_y$
for both terms of the particle wave function, the first term of the
wave function in region ``2'' with a potential barrier is similar to
the second term in the regions without potential. Therefore, the
states exactly match across the potential boundary, which should
definitely increase the total transmission amplitude. The other
possibility is that incoming angle and momentum of the second term
is totally independent of the first one and leads to resonant
transmission regardless of the transmission amplitude of the first
term. Based on our derived results, we find that the transmission
should increase even for non-split energy subband pairs and
corresponding wave function. Since the existence of certain photon
occupation numbers $N$ is determined by the laser source, we can
consider only one pair of the dressed states subbands is occupied
while the other subband pairs are unpopulated. As illustrated in
Fig.\,\ref{F3}(b), in the potential region ``2'', a particle may
populate another subband corresponding to a different number of
photons instead of changing its longitudinal momentum for the
barrier heights exactly equal to the multiple of $\hbar \omega_0$.
The opposite transition will occur at the boundary between regions
``2'' and ``3''. This will result in unimpeded tunneling $T=1$
independent of the barrier width, as seen from Fig.\,\ref{F3}(c),
which is expected to be a major contribution to the current.
\medskip

We now investigate the effect of disorder on the
transmission probability through a potential barrier in graphene.
This can appear as short-range disorder, inter-valley scattering and
trigonal distortion. In a single  layer graphene, disorder also
induces a metal-insulator transition by creating a dynamical
gap\,\cite{dis2}. Consequently, this effect can modify the gap
created by the electron-photon interaction. In bilayer grpahene,
disorder directly leads to energy dispersion with a gap as well as
modify the energy dispersion close to  the band edges. Additionally,
disorder may also lead to localized states inside a gap in bilayer
graphene\,\cite{dis3,dis1}.
\medskip

In this paper, we introduce disorder phenomenologically
through the non-conservation of the transverse electron momentum
$k_y$. As mentioned above, $k_y$ is conserved for both Dirac
electrons in infinite graphene and dressed states by photons.
Introducing the quantity $\Gamma$ as a measure for the
disorder\,\cite{yonatan,dassarma-wang}, we model its stochastic
distribution as a Lorentzian, yielding

\begin{equation}
t_{dis}(k_x,\,k_y)=\frac{\Gamma}{\pi}\int_{-\infty}^\infty dq_y\
\frac{t_0 (k_x,\,q_y) }{(k_y-q_y)^2+\Gamma^2} \ , \label{LR}
\end{equation}
where $t_0 (k_x,\,q_y)$ denotes the transmission coefficient in the
absence of any imperfections. As long as the disorder is weak with
$\Gamma \ll 1$, different distributions, which give
$\delta$-function in the limit of $\Gamma \to  0$, will result in
almost equal transmission coefficients. For nearly-head-on
collision, $k_y \ll k_{x2}$, the transmission coefficient may be
obtained analytically using a Gaussian  distribution. The imperfect
boundary of the potential region can also be the result of some
stochasticity of $k_{x2}$ to make the effect stronger. We neglect
this effect since  our goal is to investigate the role played by
disorder using a simple approximation.
\medskip

\begin{figure}[p]
\centering
\subfigure[]{
\includegraphics[height=1.7in]{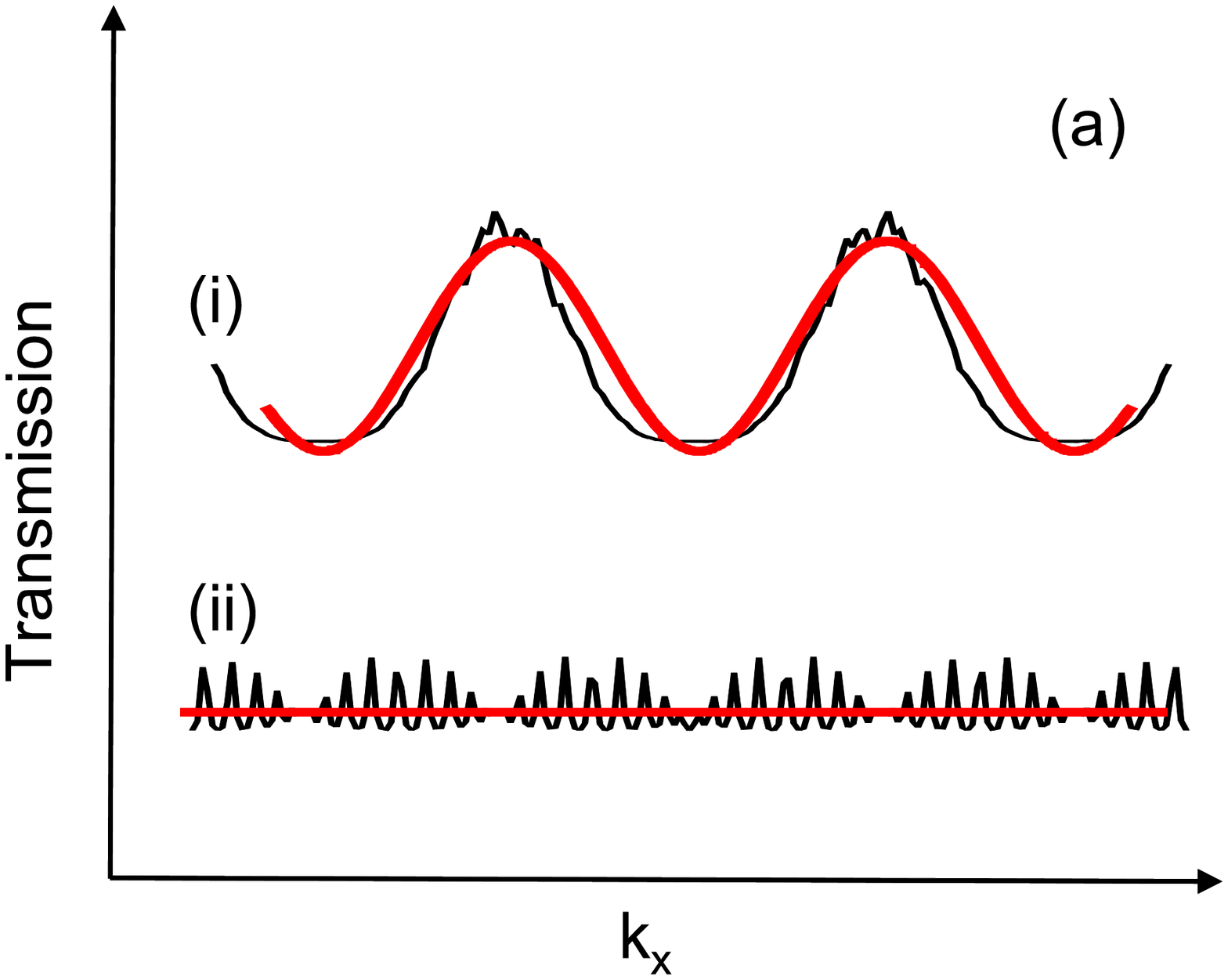}}
\subfigure[]{
\includegraphics[height=1.7in]{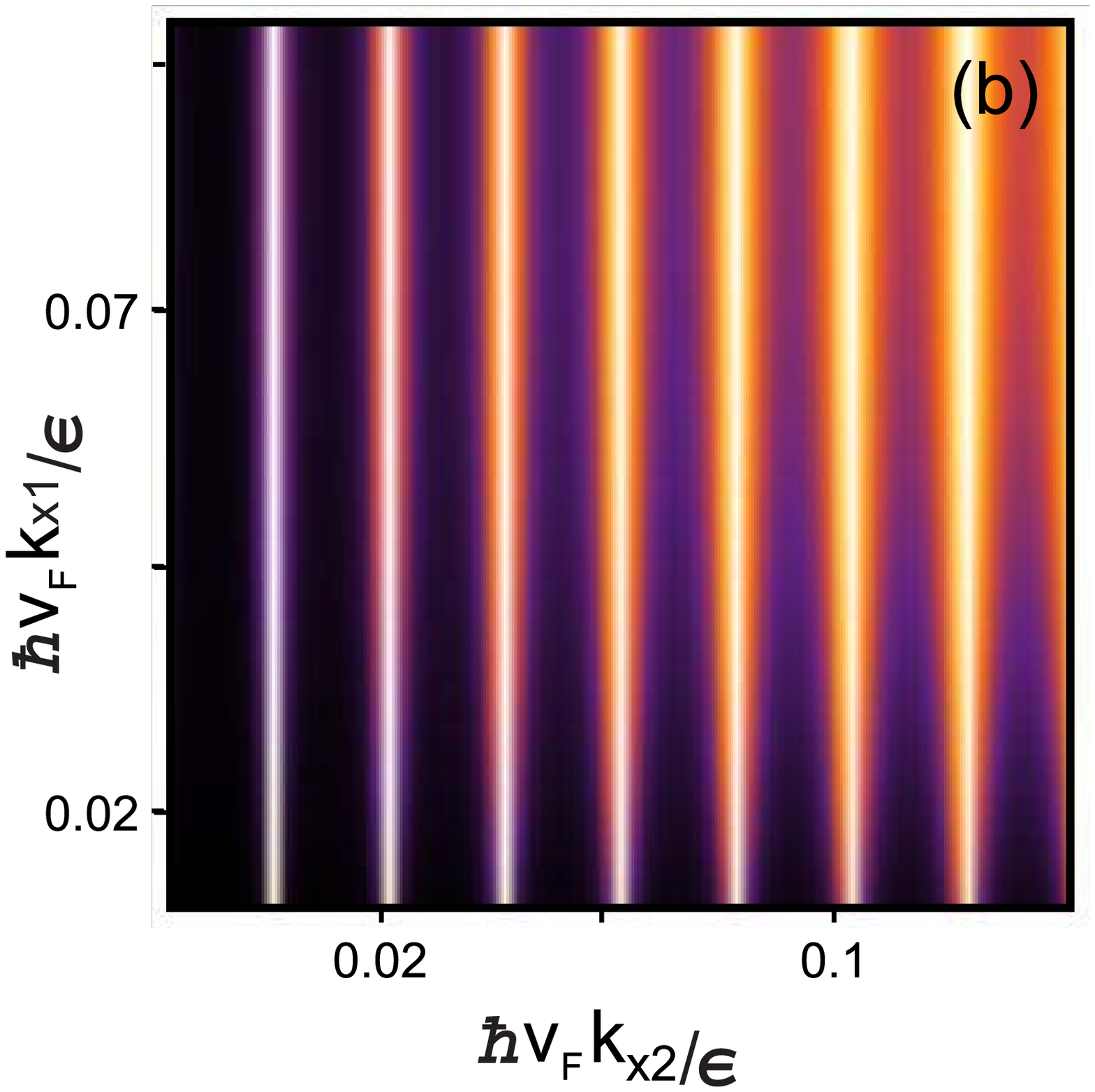}}
\subfigure[]{
\includegraphics[height=1.72in]{colorbar}}
\caption{(Color online) Effect of disorder on the transmission
probability using the Lorentzian distribution model. For the panel
(a) on the left, we plot in $(i)$ the transmission probability as a
function of $k_{x1}$ for dressed electrons (black curve) compared to
infinite graphene (red curve) for head-on collision. The plot in
$(ii)$ shows a two-peak model distribution under the influence of
disorder. For panel (b) on the right, we show the transmission
probability for electron dressed states with $\Delta=V_0/3$ in the
presence of disorder.} \label{F4}
\end{figure}

Our numerical results showing the effect due to
disorder are presented in Fig.\,\ref{F4}. First we test our
numerical results in (a) by applying the Lorentzian transformation
in \eqref{LR} to the complete transmission with $T=1$, corresponding
to the Klein paradox with the head-on collision for infinite
graphene. Analytical integration clearly results in the transmission
amplitude equal to unity. The transformed distribution demonstrates
the precision of our numerical procedure. By comparing
Fig.\,\ref{F4}(b) with Fig.\,\ref{F1}(d), we see a complete
suppression of the perfect transmission with $T=1$ along the
diagonal $k_{x1}=k_{x2}$ by disorder along the boundary as well as a
reduced range of $k_{x2}$ for photon-assisted perfect
transmission.

\section{Concluding Remarks}
\label{sec4}

According to recently published results\,\cite{Kibis}, the
interaction between Dirac particles in graphene and circularly
polarized light leads to the formation of quantum electron dressed
states. These states are appreciably different from
conventional Dirac electrons in ordinary infinite
graphene. From Ref.\,\cite{Kibis}, laser power $100$\;mW leads to
a gap on the order of $\Delta \backsimeq 100$\,meV, which is
required to make the effect significant for infrared
light frequencies and room temperature. This enables possible
experimental demonstrations of the described effects.
We have shown that electron-photon interaction gives
rise to states of broken chiral symmetry. The non-symmetrical
properties of the states become more significant when the
electron-photon interaction is increased. In addition, there are no
dressed states with chirality symmetry. In general, incoming
electrons or holes passing unimpeded through  a square potential
barrier require chiral symmetry. Under
the illumination from a circularly-polarized light, we can control
the degree of partially-broken chiral symmetry in dressed states or
the degree of partially-perfect transmission through a potential
barrier in a graphene layer.
\medskip

In the simplest approximation when only the two nearest subbands are
included, the model is formally similar to the so-called $\sigma_3$
Hamiltonian  used to describe the particles in a single layer
graphene with parabolic energy dispersion, giving non-zero electron
effective mass. We discussed the similarities as well as the
differences affecting the wave functions but not the energy
dispersion. By including more the next-nearest
subbands, the tunneling amplitude is modified in a significant
way. In the approximation when two independent pairs of subbands
are included, we obtain an enhanced transmission probability when
the barrier height is close to $\hbar\omega_0$. This is due to  the
fact that one of the terms in the corresponding wave function is
perfectly transmitted. By including more than two
independent pairs, this effect  will decrease  since perfect
transmission will occur only for the two wave function terms. The
effect is not sensitive to the barrier width, and
therefore, can be considered as a reminiscent of the Klein
paradox.
\medskip

We have introduced disorder phenomenologically in this paper through
non-conservation of the transverse electron momentum component,
which is shown to suppress the perfect transmission
along the diagonal $k_{x1}=k_{x2}$. From a physical point of view,
this disorder model could be interpreted as arising
from surface roughness of the potential barrier. The same type of
statistical distribution can be applied to
fluctuations in the the barrier width which will result in
modification of the intensity and location of the transmission
peaks. Consequently, the transmission maxima observed experimentally
will not exactly match to those theoretically
predicted for clean samples.

\section*{Acknowledgements}

This research was supported by  contract \# FA
9453-11-01-0263 of AFRL. DH would like to thank the Air Force Office
of Scientific Research (AFOSR) for its support.

\newpage

\bibliographystyle{unsrt}

\bibliography{Bibl}
\end{document}